\documentclass[10pt,a4paper]{article}

\usepackage{titling}
\title{From Function Inversion to Preimage Reporting}

\usepackage[utf8]{inputenc}
\usepackage[T1]{fontenc}
\usepackage[
a4paper,
headheight=20mm,% Set \headheight to 10mm
bottom=3cm
]{geometry}
\usepackage{layout}
\usepackage{lipsum}
\usepackage{amsfonts}
\usepackage{amsmath}
\usepackage{amssymb}
\usepackage{amstext}
\usepackage{amsthm}
\usepackage{array}

\usepackage{textcomp}
\usepackage{eurosym}

\usepackage{parskip}

\usepackage[table,dvipsnames]{xcolor}

\usepackage[shortlabels]{enumitem}
\usepackage{float}
\usepackage{graphicx}
\usepackage{mathrsfs}
\usepackage{multicol}
\usepackage{wasysym}
\usepackage{caption}
\usepackage{subcaption}
\usepackage{url}
\usepackage{csquotes}

\usepackage{algorithm}
\usepackage{algpseudocode}

\usepackage{booktabs}
\usepackage{multirow}

\usepackage{silence}
\usepackage{tikz}
\usepackage{tikz-cd}
\usetikzlibrary{arrows, backgrounds, calc, chains, decorations, patterns, positioning, shapes, arrows.meta, fit}
\usepackage{pgfplots}

\newcommand{\colorA}{blue}
\newcommand{\colorB}{Red}
\newcommand{\colorC}{ForestGreen}

\newcommand{\colorBB}{red!60}
\newcommand{\colorCB}{cyan!55!green!65}

\usepackage{pstricks}
\usepackage{pstricks-add}

\usepackage{nicematrix}

\usepackage[inkscapeformat=png]{svg}

\usepackage{mathtools, nccmath, hhline}
\usepackage{makecell}
\setcellgapes{3pt}

\usepackage[most,many,breakable]{tcolorbox}

\tcbuselibrary{theorems,skins,hooks}

\definecolor{bluebg}{HTML}{f6f9ff}
\definecolor{bluefr}{HTML}{0264A6}
\definecolor{greenbg}{HTML}{f6fffc}
\definecolor{greenfr}{HTML}{008066}
\definecolor{redbg}{HTML}{fff9f9}
\definecolor{redfr}{HTML}{ad0000}

\newtheorem{thm}{Theorem}
\newtheorem*{thm*}{Theorem}
\newtheorem{lem}[thm]{Lemma}
\newtheorem*{lem*}{Lemma}
\theoremstyle{plain}
\theoremstyle{definition}

\newtheorem*{conj*}{Conjecture}
\newtheorem{cor}[thm]{Corollary}

\newtheorem*{prob*}{Problem}

{}

\theoremstyle{remark}

\usepackage{comment}

\newcommand{\mytitle}[1]{
    \begin{center}
        \textbf{\LARGE\thetitle}
    \end{center}
    \vspace{.5cm}
}
\newcommand{\mytitlewithauthor}[1]{
    \begin{center}
        \textbf{\LARGE\thetitle}
        \\
        \theauthor
    \end{center}
    \vspace{.5cm}
}

\usepackage{fancyhdr}
\usepackage{lastpage}
\usepackage[bottom]{footmisc}
\usepackage{listingsutf8}
\usepackage{listings}
\definecolor{backgroundgray}{RGB}{250,250,250}
\definecolor{codegray}{rgb}{0.5,0.5,0.5}
\definecolor{codegray2}{rgb}{0.5,0.5,0.5}
\definecolor{codewriting}{rgb}{0.2,0.2,0.2}

\counterwithin{figure}{section}
\counterwithout{footnote}{section}
\newcommand{\authorcal}[1]{\textsc{#1}}
\newcommand{\otilde}{\tilde{O}}
\newcommand{\omegatilde}{\tilde{\Omega}}
\newcommand{\thetatilde}{\tilde{\Theta}}

\renewcommand{\Im}{\text{Im}}

\DeclareMathOperator*{\E}{E}

\usepackage[colorlinks = true,
            linkcolor = purple,
            anchorcolor = blue,
            pdftitle = {\thetitle},
            citecolor = blue]{hyperref}

\usepackage[numbers]{natbib}
\author{Mathieu Vannimmen\thanks{%
Universit\'e Libre de Bruxelles (ULB), Brussels, Belgium; \texttt{mathieu.vannimmen@ulb.be};
{\url{https://orcid.org/0009-0004-2465-855X}}{}; Mathieu Vannimmen is a Research Fellow of the Fonds de la
Recherche Scientifique - FNRS.}}

\predate{}
\postdate{}
\date{}

\begin{document}

\maketitle

\begin{abstract}
    Given an arbitrary function $f:[N]\to[M]$, the function inversion problem asks to preprocess $f$ and to be able to,
    when given some $y\in [M]$, decide whether there exists some $x\in [N]$ such that $f(x)=y$ and if so, to report
    such an $x$.
    Recently, Fiat-Naor's solution to this problem has seen significant traction in theoretical computer science, and
    allowed for several conjectures related to meta-complexity and the 3SUM problem to be broken.

    In this paper, we consider the natural extension of the function inversion problem that we call
    \emph{preimage reporting}.
    Here, instead of having to report only one $x\in [N]$ such that $f(x)=y$, the whole set $f^{-1}(y)$ must be found.

    We present a new general method to convert any arbitrary function inversion scheme into an efficient
    corresponding preimage reporting scheme.
    In particular, if $\Delta$ is the maximal indegree of $f$, we show that we can perform \emph{preimage reporting}
    using space $S$ in time $T$ for any $T$ and $S$ satisfying $TS^2=N^2 \Delta$, matching a variant of a tradeoff
    obtained by Fiat and Naor for \emph{function inversion}.
    This last result implies an improvement over the corresponding function inversion scheme itself, which constitutes
    the first function inversion scheme whose query time is inversely proportional to $|f^{-1}(y)|$.
    We derive new simple solutions for the $k$SUM-indexing with reporting and Jumbled indexing with reporting problems.
\end{abstract}

    \section{Introduction}\label{sec:introduction}

    In this section, we give the necessary background to understand our main result, and compare it with the
state of the art.

\subsection{Function inversion}\label{subsec:functioninversion}

The function inversion problem consists in preprocessing a function $f:[N]\to [N]$
into a data structure that, upon receiving a query $y\in [N]$, efficiently finds some $x\in[N]$ such that $f(x)=y$, if it
exists.
We assume that $f(x)$ is computable in time $\otilde(1)$ for all $x\in[N]$.

There are two trivial solutions to the problem.
The first one requires no preprocessing:
when given the query point $y$, we scan the domain of $f$ in time $\otilde(N)$ to find $x$ such that $f(x)=y$.
This requires no extra space.
The other approach does all the work during preprocessing: we build a hash map that associates to every $y$ an $x$ such
that $f(x)=y$, if it exists.
This approach takes time $\otilde(1)$ to answer queries but requires $\otilde(N)$ space.
Function inversion schemes try to achieve time and space between these two extremes.

The best-known result for general-purpose function inversion was given by Fiat and Naor~\citep{fiat1999}, who gave the
first tradeoff that allowed for the inversion of arbitrary functions $f:[N]\to[N]$.
They stated that for any choice of $1\leq T\leq N$ and $1\leq S\leq N$ satisfying $TS^3=N^3$, they could preprocess any
$f:[N]\to [N]$ with high probability into a data structure of size $\otilde(S)$ that answers function inversion
queries in time $\otilde(T)$ (notice that we do not require the scheme to work for values of $T$ and $S$ outside the
useful range).
We say that $TS^3=N^3$ is Fiat-Naor's tradeoff.
An equivalent statement would be to say that for any choice of $0\leq \alpha\leq 1$, they achieve function inversion
in time $T=\otilde(N^\alpha)$ and space $S=\otilde(N^{1-\alpha/3})$.
Intuitively, the more we are willing to invest into space, the faster the inversion algorithm.

In the same paper, Fiat and Naor gave an alternative tradeoff that depends on the collision probability of
$f$ over uniformly random inputs.
Recently, Kirkpatrick, Kuszmaul, Mathialagan and Vassilevska Williams~\citep{kirkpatrick2026} noticed that this very
tradeoff could be expressed in terms of $\Delta_f=\max_{y\in\Im(f)} \{|f^{-1}(y)|\}$, the maximal indegree of $f$.
They found $TS^2=N^2 \Delta_f$, which we will refer to as the \textit{$\Delta$-variant of Fiat-Naor}.

More recently, Golovnev, Guo, Peters and Stephens-Davidowitz~\citep{golovnev2023} showed that Fiat-Naor's data structure
could be modified to obtain the tradeoff $T^2 S^2=N^3$, at the cost of $\otilde(T)$ bits of shared randomness.
In the context of algorithms, the shared randomness model is often unrealistic and the tradeoff is therefore seldom
used.

While originally from cryptography, function inversion tradeoffs were adopted by other fields of computer science.
For instance, they were used to show many interesting results in complexity theory and algorithms.
In particular, such tradeoffs where used to disprove a version of the Perebor
conjecture~\citep{hirahara2024,mazor2024} and to improve the state of the art
in many preprocessed problems related to geometry~\citep{aronov2024,aronov2023,mccauley2024},
stringology~\citep{aronov2024,bibbens2026,bille2024,charalampopoulos2027,charalampopoulos2026} and
3SUM~\citep{carmel2026,dinur2026,golovnev2020,kirkpatrick2026,kopelowitz2019}.

\subsection{Main result}\label{subsec:mainresult}

A natural extension of the function inversion problem is the preimage reporting problem: instead of having to find just
one of the preimages, we need to report all of them (we need to find the set $f^{-1}(y)$).

Preimage reporting was used to solve some problems by an increasingly large set of authors.
The problem was used to attack both reporting problems~\citep{bibbens2026,bille2024,charalampopoulos2027} and
non-reporting problems~\citep{carmel2026,kirkpatrick2026,mccauley2024}.
The general idea of this last category of results is to invert a function where preimages could be either real or
false positives, and to examine all yielded values to find potential positive examples for the problem at hand
(for instance finding a pair of numbers $a$ and $b$ that sum to a particular value in the case of
preprocessed 3SUM~\citep{carmel2026,kirkpatrick2026}).

Our main result is a reduction that takes an arbitrary function inversion scheme and converts it into a corresponding
preimage reporting scheme.
Our reduction itself is greatly inspired by techniques developed by McCauley~\citep{mccauley2024} and
Kirkpatrick, Kuszmaul, Mathialagan and Vassilevska Williams~\citep{kirkpatrick2026}.
With it, we are able to convert Fiat-Naor's $TS^2=N^2 \Delta_f$ into a corresponding
preimage reporting scheme that achieves the same tradeoff $TS^2=N^2 \Delta_f$.
Note that we do so without opening the Fiat-Naor black box.

\begin{thm}\label{thm:efficientpreimagereporting}
    Let $f:[N]\to [2^w]$ be an arbitrary function to invert, with $w=\otilde(1)$.
    There exists a preimage reporting data structure that, for any choice of $T$ and $S$ satisfying $TS^2=N^2 \Delta_f$,
    achieves preimage reporting on $f$ in time $\otilde(T)$ and space $\otilde(S)$.
    Furthermore, $f$ is preprocessed in time $\otilde(N)$ with high probability.
\end{thm}

The strengths of the above result lies in its simplicity and its applicability.

Indeed, it can be used as an out-of-the-box tool that achieves state-of-the-art performance on many problems.
Examples of such uses can be found in section~\ref{subsec:otherapplications}.

Our technique does not require us to understand how function inversion schemes work as we only rely on the assumption
that they retrieve some preimage with high probability.
We deem that the sampling technique used to retrieve all preimages is natural and easy to understand.

\subsection{Comparison with previous works}\label{subsec:previousworks}

We now give existing preimage reporting results and compare our scheme to the state of the art.

Recently, Bibbens et al.~\citep[][Theorem 5]{bibbens2026} \textit{modified} the Fiat-Naor inversion scheme to make an
efficient preimage reporting scheme when $\Delta_f$ is small enough.
In particular, they showed\footnote{Strictly speaking, they showed that the data structure could be preprocessed
in expected time $\otilde(N)$, but we can achieve with high probability at a cost of multiplicative $\otilde(1)$.}:
\begin{thm}[\authorcal{Bibbens, Borevitz, McCauley}~\citep{bibbens2026}]\label{thm:preimagereportingbibbens2026}
    There exists a preimage reporting data structure that, for any choice of $T$ and $S$ satisfying $TS^3=N^3$ and
    $T=\omegatilde(\Delta_f^3)$, achieves preimage reporting in time $\otilde(T)$ and space $\otilde(S)$.
    Furthermore, $f:[N]\to[N]$ is preprocessed in time $\otilde(N)$ with high probability.
\end{thm}

Very recently, Charalampopoulos, El Ghazi, Ellert and Starikovskaya~\citep{charalampopoulos2027} combined the ideas of
Aronov et al.~\citep{aronov2024} and Bibbens
et al.~(Theorem~\ref{thm:preimagereportingbibbens2026},~\citep{bibbens2026}) to obtain a preimage reporting scheme that
covers the case where $T=\otilde(\Delta_f^3)$ that Bibbens et al.~missed.
Furthermore, they are able to use the superior tradeoff $TS^2=N^2 \Delta_f$ for this range of values of $T$.
From their result, we are able to derive the following\footnote{We slightly altered the presentation of their preimage reporting data structure to make
it fit into our $T^t S^s = N^n \Delta_f^d$ tradeoff framework.
In reality, they show that for a fixed $S$, they achieve the worst out of what $TS^3=N^3$ or $TS^2=N^2 \Delta_f$
would have given.
We can split this result in two, and obtain that they achieve $TS^3=N^3$ when $T=\omegatilde(\Delta_f^3)$ (which
    is exactly Theorem~\ref{thm:preimagereportingbibbens2026}) and $TS^2=N^2 \Delta_f$ when $T=\otilde(\Delta_f^3)$.}:
\begin{thm}[Derived from~\authorcal{Charalampopoulos, El Ghazi, Ellert, Starikovskaya}~\citep{charalampopoulos2027}]
    \label{thm:preimagereportingcharalampopoulos2027}
    There exists a preimage reporting data structure that, for any choice of $T$ and $S$ satisfying
    $TS^2=N^2 \Delta_f$ and $T=\otilde(\Delta_f^3)$, achieves preimage reporting in time $\otilde(T)$ and space
    $\otilde(S)$.
    Furthermore, $f:[N]\to[N]$ is preprocessed in time $\otilde(N)$ with high probability.
\end{thm}

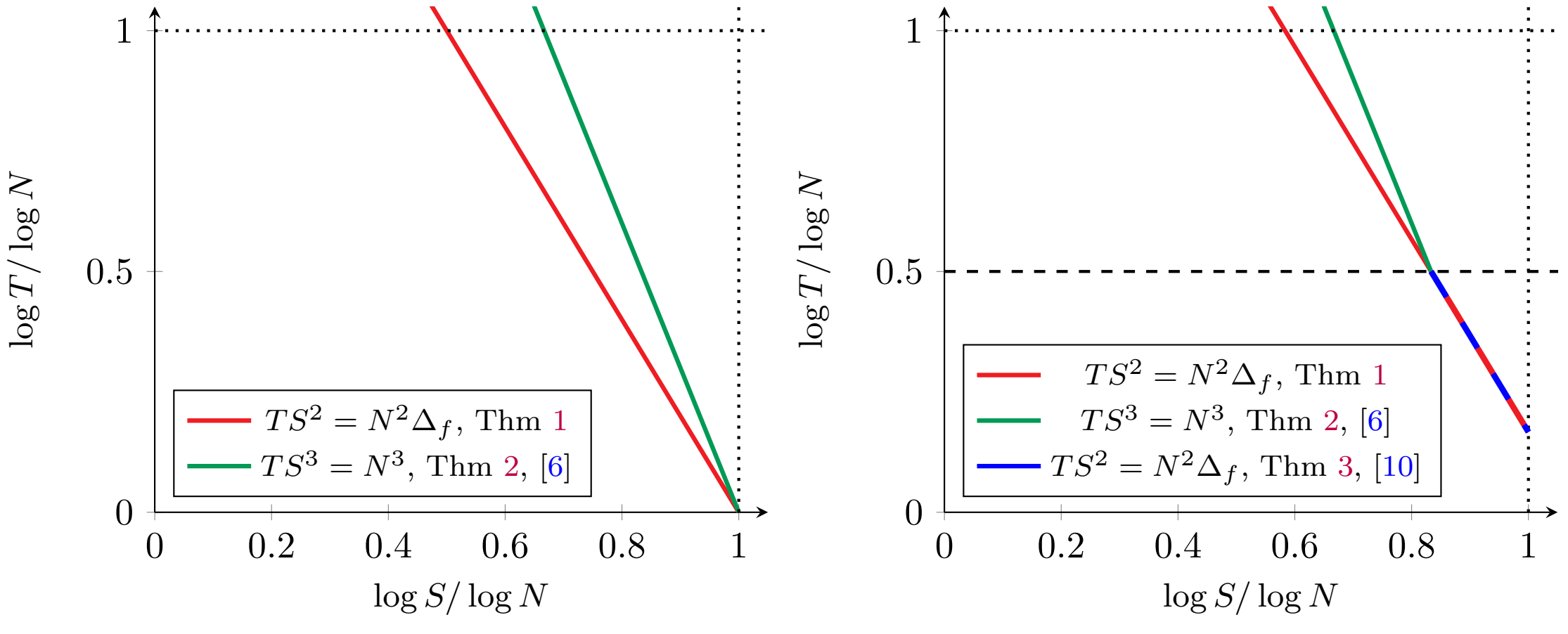
\begin{figure}[ht]
% ---------- Row 1 ----------
\begin{subfigure}[t]{0.5\textwidth}
    \centering
    \begin{tikzpicture}
\begin{axis}[
    legend style={font=\footnotesize},
    label style={font=\small},
    width=\linewidth,
    axis lines = left,
    xmin=0, xmax=1.05,
    ymin=0, ymax=1.05,
    xlabel = $\log S/\log N$,
    ylabel = $\log T/\log N$,
    enlargelimits=false,
    legend pos=south west
]
    \addplot [
    domain=0:1,
    samples=2,
    color=\colorB,
    very thick
    ]
    {2-2*x};
\addlegendentry{$TS^2=N^2 \Delta_f$, Thm~\ref{thm:efficientpreimagereporting}}
\addplot [
    domain=0:1,
    samples=2,
    color=\colorC,
    very thick,
]
{3 - 3*x};
\addlegendentry{$TS^3=N^3$, Thm~\ref{thm:preimagereportingbibbens2026},~\citep{bibbens2026}}
%Here the blue parabola is defined
    \addplot [
    domain=0:2,
    samples=2,
    dotted,
    thick
] ({1}, {x});
    \addplot [
    domain=0:2,
    samples=2,
    dotted,
    thick
] {1};

\end{axis}
\end{tikzpicture}
    \subcaption{$\Delta_f=1$}
\end{subfigure}
\hfill
\begin{subfigure}[t]{0.5\textwidth}
       \centering
    \begin{tikzpicture}
\begin{axis}[
    legend style={font=\footnotesize},
    label style={font=\small},
    width=\linewidth,
    axis lines = left,
    xmin=0, xmax=1.05,
    ymin=0, ymax=1.05,
    xlabel = $\log S/\log N$,
    ylabel = $\log T/\log N$,
    enlargelimits=false,
    legend pos=south west
]
    \addplot [
    domain=0:1,
    samples=2,
    color=\colorB,
    very thick
    ]
    {2-2*x+1/6};
\addlegendentry{$TS^2=N^2 \Delta_f$, Thm~\ref{thm:efficientpreimagereporting}}
\addplot [
    domain=0:0.8333,
    samples=2,
    color=\colorC,
    very thick,
]
{3 - 3*x};
\addlegendentry{$TS^3=N^3$, Thm~\ref{thm:preimagereportingbibbens2026},~\citep{bibbens2026}}
%Here the blue parabola is defined
\addplot [
    domain=0.8333:1,
    samples=2,
    color=\colorA,
    very thick
    ]
    {2-2*x+1/6};
\addlegendentry{$TS^2=N^2 \Delta_f$, Thm~\ref{thm:preimagereportingcharalampopoulos2027},~\citep{charalampopoulos2027}}
\addplot [
    domain=0:2,
    samples=2,
    dashed,
    thick
] {0.5};
    \addplot [
    domain=0:2,
    samples=2,
    dotted,
    thick
] ({1}, {x});
    \addplot [
    domain=0:2,
    samples=2,
    dotted,
    thick
] {1};
\addplot [
    domain=0.8333:1,
    samples=100,
    color=\colorA,
    ultra thick,
    dash pattern=on 8pt off 8pt
]
{2-2*x+1/6};

\addplot [
    domain=0.8333:1,
    samples=100,
    color=\colorB,
    ultra thick,
    dash pattern=on 8pt off 8pt,
    dash phase=8pt
]
{2-2*x+1/6};

\end{axis}
\end{tikzpicture}
    \subcaption{$\Delta_f=N^{1/6}$}
\end{subfigure}

\caption{Comparison of the query time for fixed space usage for preimage reporting schemes.}
\label{fig:comparisonplotspreimagereporting}
\end{figure}

We can now compare Theorems~\ref{thm:efficientpreimagereporting},~\ref{thm:preimagereportingbibbens2026}
and~\ref{thm:preimagereportingcharalampopoulos2027}.
Given only the exponents $t,s,n,d$ in the tradeoff $T^t S^s =N^n \Delta_f^d$, it is not clear which are better.
To compare them, we can fix $S=N^\gamma$ for some $0\leq \gamma\leq 1$ and get the corresponding value of $T$ back.
In figure~\ref{fig:comparisonplotspreimagereporting}, we did so and plotted the exponents for the three schemes for two
different values of $\Delta_f$.

In particular, our result gives a simple tradeoff that is at least as good as the state of the art for any choice of
$S$ (or $T$).

Section~\ref{subsec:preimagereporting} gives all the ideas used to prove Theorem~\ref{thm:efficientpreimagereporting}.
The formal proofs and the precise reduction can be found in section~\ref{subsec:preimagereportingreduction}.
Section~\ref{sec:applications} gives examples of applications, in which we improve the state of the art for some
reporting problems.

    \section{Efficient preimage reporting}\label{sec:efficientpreimagereporting}

    \subsection{Preliminaries}\label{subsec:definitions}

We denote by $[N]$ the set $\{1,\ldots,N\}$, and we use $\otilde$ notation which omits polylogarithmic factors, such
that $\otilde(g(n))=O(g(n))\log^{O(1)} n$ (we define $\thetatilde$ and $\omegatilde$ analogously).
Let $g:D\to C$ be some function and let $D'\subseteq D$.
We define $D$ to be the \textit{domain} of $g$, $C$ to be its \textit{codomain}, $g(D')=\{g(x)\mid x\in D'\}$ to denote
the image of a set $D'$ and $\Im(g)=g(D)\subseteq C$ to be the image of $g$.
The inverse of $y\in C$ by $g$ is defined to be $g^{-1}(y)=\{x\in D\mid g(x)=y\}$.
We use $\log$ to denote $\log_2$.
We work in the word RAM model, with words of size $\Omega(\log N)$.

A \textit{function inversion data structure achieves function inversion in time $T$ and space $S$} if it preprocesses
any given function $f:[N]\to[N]$ into a data structure of space $S$ and,
given a query point $y\in [N]$, it finds a point in $x\in [N]$ such that $f(x)=y$ (or says that it does
not exist) in time $T$.
We define preimage reporting data structures equivalently.

In general, tradeoffs take the following form $T^t S^s=N^n \Delta_f^d$, for $t,s,n>0$, $d\geq 0$ constants.
We say that a \textit{function inversion scheme achieves the tradeoff} $T^t S^s =N^n \Delta_f^d$ if, for any choice of
$T$ and $S$ satisfying $T^t S^s =N^n \Delta_f^d$, $1\leq T\leq N$ and $1\leq S\leq N$, there exists a function inversion data structure
that achieves function inversion in time $\otilde(T)$ and space $\otilde(S)$.
We define preimage reporting schemes equivalently.

The above allows us to state Fiat-Naor's and Golovnev et al.'s theorems properly.

\begin{thm}[\authorcal{Fiat-Naor}~\citep{fiat1999}]\label{thm:fiatnaor}
    There exists a function inversion scheme that achieves the tradeoff $TS^3=N^3$.
    Furthermore, $f:[N]\to[N]$ is preprocessed in time $\otilde(N)$ with high probability.
\end{thm}

\begin{thm}[$\Delta$-variant of \authorcal{Fiat-Naor}~\citep{fiat1999,kirkpatrick2026}]\label{thm:fiatnaordelta}
    There exists a function inversion scheme that achieves the tradeoff $TS^2=N^2 \Delta_f$.
    Furthermore, $f:[N]\to[N]$ is preprocessed in time $\otilde(N)$ with high probability.
\end{thm}

\begin{thm}[\authorcal{Golovnev, Guo, Peters, Stephens-Davidowitz}~\citep{golovnev2023}]\label{thm:golovnev}
    There exists a function inversion scheme that achieves the tradeoff $T^2 S^2=N^3$.
    Furthermore, $f:[N]\to[N]$ is preprocessed in time $\otilde(N)$ with high probability.
    The scheme uses $\otilde(T)$ words of shared randomness.
\end{thm}

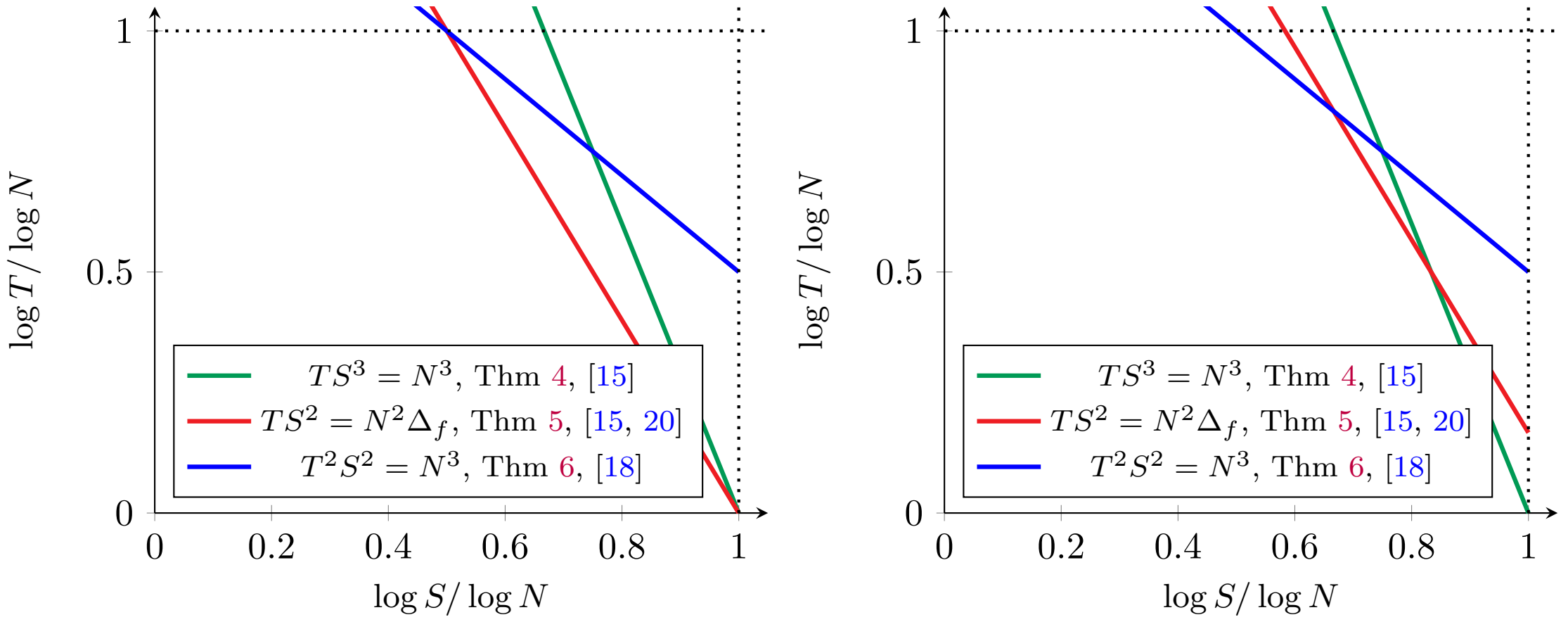
\begin{figure}[ht]
% ---------- Row 1 ----------
\begin{subfigure}[t]{0.5\textwidth}
    \centering
    \begin{tikzpicture}
\begin{axis}[
    legend style={font=\footnotesize},
    label style={font=\small},
    width=\linewidth,
    axis lines = left,
    xmin=0, xmax=1.05,
    ymin=0, ymax=1.05,
    xlabel = $\log S/\log N$,
    ylabel = $\log T/\log N$,
    enlargelimits=false,
    legend pos=south west
]
\addplot [
    domain=0:1,
    samples=2,
    color=\colorC,
    very thick,
]
{3 - 3 * x};
\addlegendentry{$TS^3=N^3$, Thm~\ref{thm:fiatnaor},~\citep{fiat1999}}
\addplot [
    domain=0:1,
    samples=2,
    color=\colorB,
    very thick
    ]
    {2-2 * x};
\addlegendentry{$TS^2=N^2 \Delta_f$, Thm~\ref{thm:fiatnaordelta},~\citep{fiat1999,kirkpatrick2026}}
\addplot [
    domain=0:1,
    samples=2,
    color=\colorA,
    very thick
    ]
    {3/2-x};
\addlegendentry{$T^2 S^2=N^3$, Thm~\ref{thm:golovnev},~\citep{golovnev2023}}
    \addplot [
    domain=0:2,
    samples=2,
    dotted,
    thick
] ({1}, {x});
    \addplot [
    domain=0:2,
    samples=2,
    dotted,
    thick
] {1};

\end{axis}
\end{tikzpicture}
    \subcaption{$\Delta_f=1$}
\end{subfigure}
\hfill
\begin{subfigure}[t]{0.5\textwidth}
       \centering
    \begin{tikzpicture}
\begin{axis}[
    legend style={font=\footnotesize},
    label style={font=\small},
    width=\linewidth,
    axis lines = left,
    xmin=0, xmax=1.05,
    ymin=0, ymax=1.05,
    xlabel = $\log S/\log N$,
    ylabel = $\log T/\log N$,
    enlargelimits=false,
    legend pos=south west
]
\addplot [
    domain=0:1,
    samples=2,
    color=\colorC,
    very thick,
]
{3 - 3*x};
\addlegendentry{$TS^3=N^3$, Thm~\ref{thm:fiatnaor},~\citep{fiat1999}}
%Here the blue parabola is defined
\addplot [
    domain=0:1,
    samples=2,
    color=\colorB,
    very thick
    ]
    {2-2*x+1/6};
\addlegendentry{$TS^2=N^2 \Delta_f$, Thm~\ref{thm:fiatnaordelta},~\citep{fiat1999,kirkpatrick2026}}
\addplot [
    domain=0:1,
    samples=2,
    color=\colorA,
    very thick
    ]
    {3/2-x};
\addlegendentry{$T^2 S^2=N^3$, Thm~\ref{thm:golovnev},~\citep{golovnev2023}}
    \addplot [
    domain=0:2,
    samples=2,
    dotted,
    thick
] ({1}, {x});
    \addplot [
    domain=0:2,
    samples=2,
    dotted,
    thick
] {1};

\end{axis}
\end{tikzpicture}
    \subcaption{$\Delta_f=N^{1/6}$}
\end{subfigure}

\caption{Comparison of the query time for fixed space usage for function inversion schemes.}
\label{fig:comparisonplotsinversion}
\end{figure}

In general, we might be interested in inverting functions where the codomain is larger than the domain:
functions of the form $f:[N]\to[M]$ with $M\geq N$.
Furthermore, we let $M=2^{\otilde(1)}$ to make it fit into $\otilde(1)$ machine words.
We call inversion schemes that are able to invert such functions \textit{extended} inversion schemes.

It is by now well known that classical inversion schemes whose tradeoffs do not depend on $\Delta_f$
(where $d=0$ in $T^t S^s=N^n \Delta_f^d$,
such as e.g.~$TS^3=N^3$) can be turned into corresponding extended inversion schemes via standard
hashing techniques (see e.g.~\citep{golovnev2020,kirkpatrick2026,kopelowitz2019}).
The original idea essentially works by inverting in $h\circ f$ instead of $f$, where $h$ is picked uniformly at random
from a $2$-universal hash family.
The issue when there is a dependency on $\Delta_f$ ($d>0$) is that this hash function might introduce large
discrepancies between $\Delta_{h\circ f}$ and $\Delta_f$.
Picking $h$ from a $\Theta(\log N)$-wise independent hash family gives $\Delta_{h\circ f}=\otilde(\Delta_f)$ using
standard balls and bins results (e.g.~\citep{celis2011}).
The full proof is given for completeness in appendix~\ref{sec:codomainextension}.

\begin{thm}[Codomain extension]\label{thm:codomainextension}
    Let $t,s,n>0$, $d\geq 0$ be constants.
    If there exists a function inversion data structure that achieves the tradeoff $T^t S^s=N^n \Delta_f^d$ that
    preprocesses $f:[N]\to[N]$ in time $P$, then there exists a
    corresponding extended function inversion data structure that achieves the same tradeoff and preprocesses
    $f:[N]\to[M]$ for $M=2^{\otilde(1)}$ in time $\otilde(P)$ with high probability.
\end{thm}

\subsection{The main idea}\label{subsec:preimagereporting}

We now give the basis for our method, which is inspired by the works of McCauley~\citep{mccauley2024} in 2024 and
Kirkpatrick, Kuszmaul, Mathialagan and Vassilevska Williams~\citep{kirkpatrick2026} in 2026.
The idea is as follows: instead of inverting only one function $f:[N]\to[M]$ of maximal indegree $\Delta_f$, build a
partition of the domain $[N]$ into $K=\thetatilde(\Delta_f)$ sets $D_i$ for $1\leq i\leq K$, where each element of
$[N]$ is placed in $D_i$ with probability $1/K$ (in particular, we have $|D_i|=\Theta(N/K)$ on expectation).
Then, we define functions $f_i:D_i\to[M]:x\mapsto f(x)$ for all $1\leq i\leq K$.
For each $f_i$, we can preprocess a function inversion data structure.
We can show that the number of preimages of any given $y$ by any $f_i$ will be $O(\log \Delta_f)=O(\log N)$.
We can even expect that with high enough probability, there will be exactly one preimage of $y$ by any given $f_i$ and
inverting in each of the $K=\otilde(\Delta_f)$ subfunctions $f_i$ will yield a large enough fraction of the at most
$\Delta_f$ preimages of the query point.
Repeating $\otilde(1)$ times the above scheme would then enable us to find $f^{-1}(y)$ for any $y$ with high
probability.

This idea uses function inversion schemes as a black box to perform preimage reporting, which is an advantage.
However, the main flaw of the approach is the partition of $[N]$ into $\otilde(\Delta_f)$ sets: as they are defined via
a random function, they require $\Theta(N)$ space to be stored.
This was mitigated in two different ways:
\begin{itemize}
    \item McCauley~\citep{mccauley2024} figured that the same partition could be used for at most $N$ different
    functions.
    In particular, if one needs to do preimage reporting on multiple functions, the overhead of storing a random
    partition of $[N]$ could be amortized.
    \item Kirkpatrick, Kuszmaul, Mathialagan and Vassilevska Williams~\citep{kirkpatrick2026} noticed that
    if the function that they were trying to invert was of a special
    form\footnote{$f:[\Delta_f]\times [N/\Delta_f]\to [M], \forall y\in [M]: (a,b)\neq (a',b')
    \in f^{-1}(y) \Longrightarrow a\neq a'\land b\neq b'$}, they could get away with only partitioning a subset of
    $[N]$.
\end{itemize}

The main idea of our work is to use a hash function $h$ picked uniformly at random from a $\Theta(\log N)$-wise
independent hash family to do the partitioning.
Then, instead of considering $f$, we consider $f'=f\circ h$.
Given some partition of $[N]$ into $K$ sets $D_i$ for $1\leq i\leq K$ of size $|D_i|=\thetatilde(N/\Delta_f)$,
we define the subfunctions $f_i:D_i\to [M]:x\mapsto f(h(x))$.
We will show in section~\ref{subsec:preimagereportingreduction} that:
\begin{enumerate}
    \item With constant probability and for a fixed $x\in [N]$, there is some $i$ such that $x$ is the only preimage
    of $y$ by $f$ that can be found through $f_i$ (there is an $i$ such that $h(D_i)\cap f^{-1}(y)=\{x\}$).
    \item $\Delta_{f_i}=\otilde(1)$ for all $i$ with high probability.
\end{enumerate}

Replicating the construction an appropriate number of times, we find the following lemma.

\begin{figure}[ht]
    \centering
    \captionsetup{justification=centering}
    \begin{tikzpicture}
    \begin{scope}[every node/.style={circle,thick,draw,minimum size=4mm}]
        \node (y) [fill=\colorBB] at (4,2) {$y$};
        \node[draw=none] (f) at (2.5, 4) {$f$};
        \node[draw=none] (h) at (-0.5, 4) {$h$};
        \node (x1) at (1,3) {$x_1$};
        \node[draw=none] (x2) at (1,2) {\raisebox{-4.5pt}[0pt][0pt]{$\vdots$}};
        \node (x3) at (1,1) {$x_{\Delta}$};
        \node (v1) at (1,-1) {$v_1$};
        \node[draw=none] (v2) at (1,-2) {\raisebox{-4.5pt}[0pt][0pt]{$\vdots$}};
        \node (v3) at (1,-3) {$v_n$};
        \node (w1) at (4,-1) {$w_1$};
        \node[draw=none] (w2) at (4,-2) {\raisebox{-4.5pt}[0pt][0pt]{$\vdots$}};
        \node (w3) at (4,-3) {$w_{m}$};
        \node (s1) [fill=\colorCB] at (-2,4) {$a_1$};
        \node (s2) [fill=\colorCB] at (-2,3) {$a_2$};
        \node[draw=none] (s3) at (-2,2) {\raisebox{-4.5pt}[0pt][0pt]{$\vdots$}};
        \node (s4) at (-2,1) {$a_{b}$};
        \node (u1) [fill=\colorCB] at (-2,-1) {$u_1$};
        \node[draw=none] (u2) at (-2,-2) {\raisebox{-4.5pt}[0pt][0pt]{$\vdots$}};
        \node (u3) at (-2,-3) {$u_{\ell}$};
    \end{scope}

    \node[draw, rounded corners, fit=(x1) (x2) (x3)] (X) {};
    \node[draw=none,above=3pt of X] {$f^{-1}(y)$};
    \node[draw, rounded corners, fit=(s1) (s2) (s3) (s4)] (S) {};
    \node[draw=none,above=3pt of S] {$D_i$};
    \node[draw, rounded corners, fit=(u1) (u2) (u3)] (U) {};
    \node[draw=none,below=3pt of U] {$[N]\setminus D_i$};
    \node[draw, rounded corners, fit=(v1) (v2) (v3)] (V) {};
    \node[draw=none,below=3pt of V] {$[N]\setminus f^{-1}(y)$};
    \node[draw, rounded corners, fit=(w1) (w2) (w3)] (W) {};
    \node[draw=none,below=3pt of W] {$[M]\setminus \{y\}$};

    \begin{scope}[>={Stealth[width=6pt, length=6pt]},
                  every node/.style={fill=white,circle},
                  every edge/.style={draw=black,very thick}]
        \path [->] (x1) edge (y);
        \path [->] (x3) edge (y);
        \path [->] (v1) edge (W);
        \path [->] (v3) edge (W);
        \path [->] (u1) edge (X);
        \path [->] (u3) edge (V);
        \path [->] (s1) edge (X);
        \path [->] (s2) edge (X);
        \path [->] (s4) edge (V);
    \end{scope}
    \end{tikzpicture}
    \caption{A visualization of $f'$, for some fixed $y\in [M]$ and $D_i$.
    The set $f'^{-1}(y)$ is depicted in green.}
    \label{fig:visualizationfprime}
\end{figure}
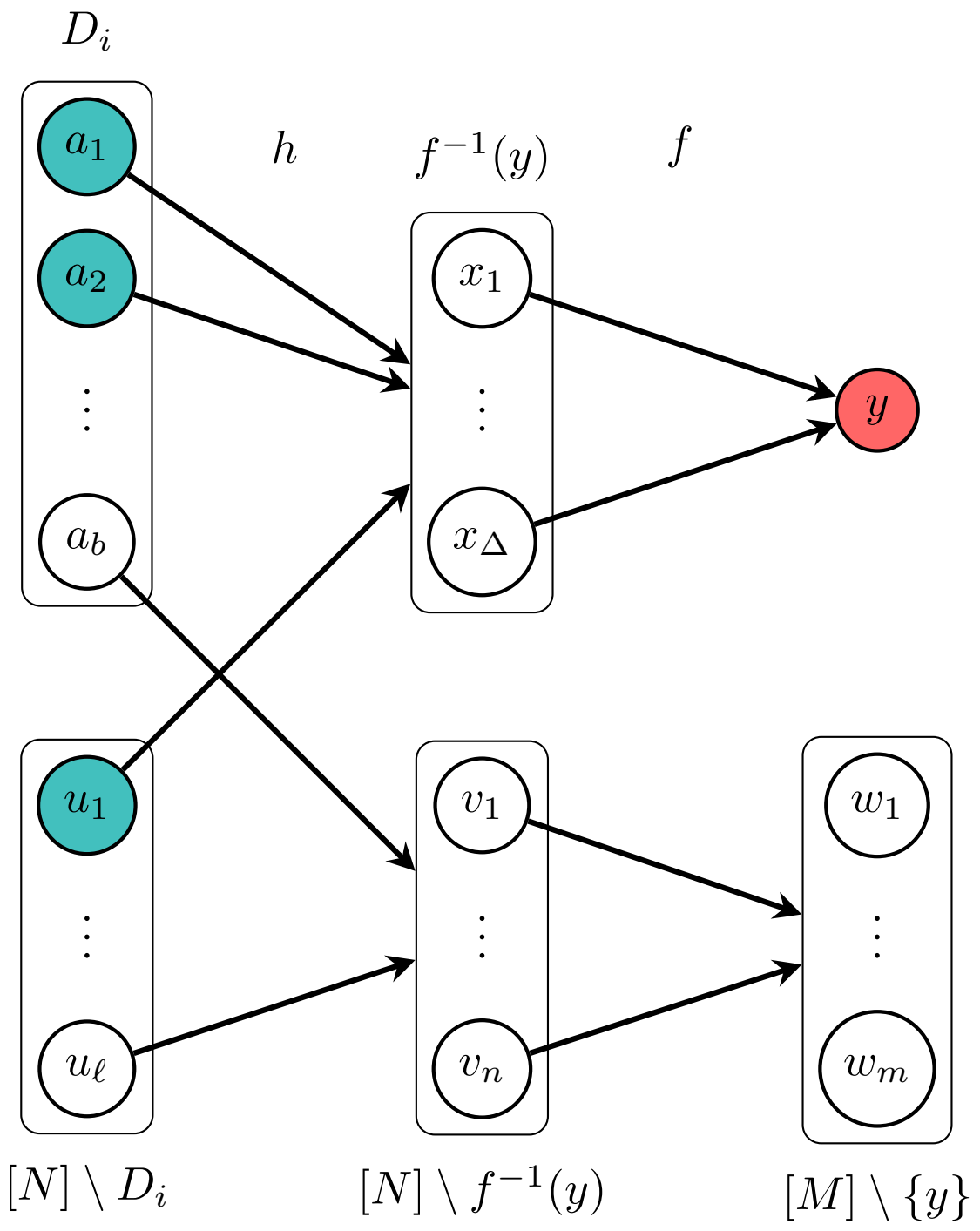

\begin{lem}\label{lem:improvedfullinversion}
    Let $t,s,n,p>0$, $d\geq 0$ be constants.
    If there exists an inversion scheme that achieves the tradeoff $T^t S^s =N^n \Delta_f^d$ and preprocesses
    $f:[N]\to[N]$ in time $\otilde(N^p)$, then there exists a
    corresponding extended preimage reporting data structure that preprocesses $f:[N]\to[M]$ with $M=2^{\otilde(1)}$ in
    time $\otilde(\Delta_f^{1-p}N^p)$ with high probability.
    Furthermore, for all choices of $T$ and $S$ satisfying $T^t S^s =N^n \Delta_f^{t+s-n}$,
    the data structure requires space $\otilde(S)$ and finds the set $f^{-1}(y)$ of any query point $y\in [M]$ in time
    $\otilde(T)$.
\end{lem}

Applying the above result to the $\Delta$-variant of Fiat-Naor (Theorem~\ref{thm:fiatnaordelta}, $t=d=p=1,s=n=2$)
yields Theorem~\ref{thm:efficientpreimagereporting}.
One can check that plugging in the two other inversion schemes (Theorems~\ref{thm:fiatnaor} and~\ref{thm:golovnev})
yield strictly worse preimage reporting data structures (for any fixed $S=N^{\gamma}$, the corresponding $T$ is
larger for these two schemes).

\subsection{Tools}\label{subsec:tools}

\begin{lem}[{\authorcal{Bellare, Rompel}~\citep[][Lemma 2.3]{bellare1994}}]\label{lem:kwiseconcentrationbound}
    Let $k\geq 4$ be an even integer.
    Suppose $X_1,\ldots,X_n$ are $k$-wise independent random variables taking values in $[0,1]$.
    Let $X=X_1+\cdots+X_n$ and $\mu=\E[X]$, and let $T>0$.
    Then, \[\Pr[|X-\mu|\geq T]\leq 8\cdot \left(\frac{k\mu+k^2}{T^2}\right)^{k/2}\]
\end{lem}

\begin{thm}[{\authorcal{Yao}~\citep{yao1990}}]\label{thm:yao1990}
    Any function inversion scheme with a success probability $\Omega(1)$ that uses $S$ words of space and answers queries
in time $T$ must satisfy:
    \[TS=\omegatilde(N)\]
\end{thm}

\begin{lem}\label{lem:tradeoffexponents}
Let $t,s,n>0$, $d\geq 0$ be constants and
    suppose there exists a function inversion scheme that achieves the tradeoff $T^t S^s = N^n \Delta_f^d$.
    Then, we must have:
    \[n\geq s\hspace*{3em}n\geq t\]
\end{lem}

\begin{proof}
    By our definition of inversion schemes, the scheme must achieve the tradeoff $T^t S^s=N^n \Delta_f^d$ for all
    possible choices of $1\leq T,S\leq N$ satisfying said tradeoff, and so for any value of $\Delta_f$.
    If $\Delta_f=1$ and if we fix $T=N^\alpha$, we obtain $S=N^{(n-\alpha t)/s}$.
    Since $1\leq T,S\leq N$, we obtain that the function inversion scheme must be instantiable for all values of
    $\alpha$ bounded by $\max\{0,(n-s)/t\}\leq \alpha\leq \min\{1,n/t\}$.
    We will show that $n\geq s$ and $n\geq t$ via a contradiction with Yao's Theorem~\ref{thm:yao1990}.

    We have $T^t S^s = N^n \Delta^d \Longleftrightarrow S=N^{n/s}\Delta_f^{d/s} T^{-t/s}\Longleftrightarrow TS=N^{n/s}\Delta_f^{d/s} T^{1-t/s}$.
    Let us first show that $n\geq s$.
    We fix $\Delta_f=1$.
    We know from our reasoning above that picking $T=N^\alpha$ with $\alpha=\max\{0,(n-s)/t\}$ gives a valid point
    on the tradeoff curve.
    We distinguish two cases: either $(n-s)/t\geq 0$, which implies that $n\geq s$ and we are done, or
    $(n-s)/t\leq 0$, such that $\max\{0,(n-s)/t\}=0$.
    If this is the case, we have $TS=N^{n/s+(\max\{0,(n-s)/t\})(1-t/s)}=N^{n/s}$ and $N^{n/s}=\omegatilde(N)$ from
    Theorem~\ref{thm:yao1990}, which immediately gives
    $n/s\geq 1\Longleftrightarrow n\geq s $, which contradicts our hypothesis that $(n-s)/t\leq 0$.
    Hence, $n\geq s$.

    Repeating the same argument with $\alpha=\min\{1,n/t\}$ yields $n\geq t$.
\end{proof}

\subsection{A preimage reporting reduction}\label{subsec:preimagereportingreduction}

Unless stated otherwise, we use the random polynomial hash families of Carter and Wegman~\citep{wegman1980}, where each
hash function is a polynomial of degree $k-1$ and the family is $k$-wise independent.

In this section, we complete the sketch of section~\ref{subsec:preimagereporting} and show that the ideas do work.

Let $t,s,n>0$, $d\geq 0$ be constants.
We assume access to an extended (or not, using Theorem~\ref{thm:codomainextension}) function inversion scheme
that achieves the tradeoff $T^t S^s=N^n \Delta_f^d$.
Consider a function $f:[N]\to[M]$ such that $M=2^{\otilde(1)}$, that we want to invert.
Instead of inverting in $f$ directly, we will invert in $f':[N]\rightarrow [M]:x\mapsto f(h(x))$ for $h:[N]\to [N]$
a hash function picked uniformly at random from a $k$-wise independent family of hash functions $\mathcal{H}$, with
$k=\lceil \log N\rceil$.
Consider a partition of $[N]$ into
$K=\Delta_f \lceil\log N \rceil$ disjoint sets denoted $D_i$, each of size $|D_i|=\Theta(N/K)$.
We can then define $K$ functions $f_i: D_i\to [M]:x\mapsto f'(x)$ which are just
domain-restricted versions of $f'$.
The main idea is that if we fix some query $y$, we expect that in each set $D_i$, we will have at most
$\otilde(|f^{-1}(y)|/K)=\otilde\left(|f^{-1}(y)|/\Delta_f\right)=\otilde(1)$ preimages of $y$ by $f'$, because the domain was uniformly
shuffled and the preimages distributed uniformly in the sets $D_i$.
Note that $h$ can be both queried and stored in time and space $O(\log N)=\otilde(1)$ and
the overhead of using such an $h$ vanishes in the $\otilde$ notation.

Notice that if $x\in f_i^{-1}(y)$, then $h(x)$ is a preimage of $y$ by $f$.

The main intuition of this whole section is that we want to shuffle the domain of $f$.
Assuming the shuffle is good enough, dividing the domain in $\otilde(\Delta_f)$ pieces
yields sets $D_i$ where only $\otilde(1)$ (or ideally just $1$) preimages of $y$ can be found, meaning that we have
$\otilde(\Delta_f)$ functions $f_i$ with domains of size $\thetatilde(N/\Delta_f)$ and $\Delta_{f_i}=\otilde(1)$.

The preimage reporting algorithm follows.
\begin{enumerate}
    \item Pick $h$ from a $\lceil \log N\rceil$-wise independent hash family
    and choose a value of $T$ and $S$ on the curve $T^t S^s=|D_i|^n \Delta_{f_i}^d=\otilde(N^n/\Delta_f^n)$.
    Choose a partition of $[N]$ into $K=\Delta_f \lceil\log N\rceil$ sets $D_i$ of size $\thetatilde(N/\Delta_f)$.
    \item Preprocess $K$ independent function inversion schemes on the functions
    $f_i:D_i\to [M]:x\mapsto f(h(x))$ with the chosen parameters $T$ and $S$.
    \item When given a query point $y\in [M]$, query all $K$ inversion schemes.
    Let $x_i$ be the output of the inversion of $f_i$ on $y$.
    Then, $h(x_i)\in f^{-1}(y)$.
    Output all the preimages of $y$ that you find.
\end{enumerate}
To obtain a high probability of success, one can replicate this construction $\log^2 N$ times and output the union of the
outputs.

To show that this all works, we first need to show that the maximal indegree $\Delta_{f_i}$ of the functions $f_i$ is
$\otilde(1)$.

\begin{lem}\label{lem:hasheddomaindelta}
    With high probability, $\Delta_{f_i}=\otilde(1)$ holds for all $f_i$ simultaneously.
\end{lem}

\begin{proof}
    Let us fix a function $f_i:D_i\to [M]$ and a $y\in \Im(f)$.

    Let us define the binary random variable $X_{i,d}$ that is 1 if and only if $d\in f_i^{-1}(y)$,
    such that $X_i=\sum_{d\in D_i}X_{i,d}$ denotes the size of $f_i^{-1}(y)$, the number of preimages of $y$ by $f_i$.

    For a fixed $f_i$, the random variables $X_{i,d}$ are $k$-wise independent, since the randomness comes from the
    choice of $h$.
    To prove our lemma, we would like to apply Lemma~\ref{lem:kwiseconcentrationbound} which is a version of Chernoff
    bounds for $k$-wise independent random variables (which requires $k$ to be even, we can consider $k-1$ otherwise).
    We have $\E[X_i]=\mu=|f^{-1}(y)|/(\Delta_f \log N)\leq 1$ by linearity of expectation and:
    \begin{align*}
        \Pr\left[X_i\geq k^2+1\right]
        &\leq\Pr\left[\left|X_i-\mu\right|\geq k^2\right]
        \leq 8\cdot\left(\frac{k+k^2}{k^4}\right)^{k/2}
        \leq 8\cdot\left(\frac{1}{k^3}+\frac{1}{k^2}\right)^{k/2}
    \end{align*}

    Since $k=\Theta(\log N)$, we have that $X_i<k^2+1=\otilde(1)$ with probability at least $1-N^{-\omega(1)}$.
    Union bounding over all $\Theta(\Delta_f\log N)$ functions $f_i$ and all $O(N)$ possible values of $y$ in $\Im(f)$ gives
    the result.
\end{proof}

We now show that a constant fraction of the preimages of $y$ can be retrieved using classical
function inversion even though we are hashing the domain.

\begin{lem}\label{lem:hasheddomainonepreimage}
    Consider a fixed $y\in [M]$ and a fixed $x\in f^{-1}(y)$.
    Then, there exists at least one set $D_i$ such that $h(D_i)\cap f^{-1}(y)=\{x\}$ with probability at least
    $1/3$.
\end{lem}

\begin{proof}
    Let $A_d$ be the event $\{h(d)=x\}$ for $d\in [N]$ and let $A=\bigcup_{d\in [N]}A_d$ (there is a way to reach $x$ using $h$).
    Let $B_{i,d,d'}$ be the event $\{h(d)=x\land h(d')\in f^{-1}(y)\}$ for $1\leq i\leq K$, $d,d'\in D_i$
    and let $B=\bigcup_{i\in [K]}\bigcup_{d\neq d'\in D_i} B_{i,d,d'}$ (there is some set $D_i$ from which $x$ is reachable and is not the
    only element of $f^{-1}(y)$ to be so).
    The lemma requires a bound on $\Pr[A\setminus B]=\Pr[A]-\Pr[A\cap B]\geq \Pr[A]-\Pr[B]$.
    We have, by Bonferroni inequalities and by pairwise independence of $h$:
    \[\Pr[A]\geq \sum_{d\in [N]}\Pr[A_d]-\sum_{d< d'\in [N]}\Pr[A_d\cap A_{d'}]=
    N\cdot \frac{1}{N}-\binom{N}{2}\frac{1}{N^2}=\frac{N+1}{2N}\geq \frac{1}{2}\]
    By union bound and by pairwise independence of $h$:
    \[\Pr[B]\leq \Theta\left(K \cdot\frac{N^2}{K^2}\cdot\frac{1}{N}\cdot\frac{|f^{-1}(y)|}{N}\right)\leq
    \Theta\left(\frac{|f^{-1}(y)|}{K}\right)\leq \Theta\left(\frac{\Delta_f}{\Delta_f \log N}\right)=
    \Theta\left(\frac{1}{\log N}\right)\]
    For large enough $N$, $\Pr[A]-\Pr[B]\geq 1/2-\Theta(1/\log N)\geq 1/3$.
\end{proof}

We are now ready to show Lemma~\ref{lem:improvedfullinversion}, that we restate.
\begin{lem*}
    Let $t,s,n,p>0$, $d\geq 0$ be constants.
    If there exists an inversion scheme that achieves the tradeoff $T^t S^s =N^n \Delta_f^d$ and preprocesses
    $f:[N]\to[N]$ in time $\otilde(N^p)$, then there exists a
    corresponding extended preimage reporting data structure that preprocesses $f:[N]\to[M]$ with $M=2^{\otilde(1)}$ in
    time $\otilde(\Delta_f^{1-p}N^p)$ with high probability.
    Furthermore, for all choices of $T$ and $S$ satisfying $T^t S^s =N^n \Delta_f^{t+s-n}$,
    the data structure requires space $\otilde(S)$ and finds the set $f^{-1}(y)$ of any query point $y\in [M]$ in time
    $\otilde(T)$.
\end{lem*}

\begin{proof}
    Let us use the construction described at the beginning of this section to obtain a new
    function $f':[N]\to [M]$, and the associated $f_i:D_i \to [M]:x\mapsto f'(x)$ functions
    (we have $\Theta(\Delta_f \log N)$ of these), with $|D_i|=\Theta(N/\Delta_f\log N)$.
    By Lemma~\ref{lem:hasheddomaindelta}, $\Delta_{f_i}=\otilde(1)$ with high probability for all $i$.
    Let $T_i$ and $S_i$ be respectively the query time and the space of the function inversion scheme for subfunction
    $f_i$.
    We choose the following values for $T_i$ and $S_i$, that only depend on a single parameter $\alpha$:
    \[T_i=N^{\alpha}\hspace*{3em}
    S_i=N^{(n-\alpha t)/s}/\Delta_f^{n/s}\]
    One can easily check that the function inversion schemes's tradeoff is respected
    ($T_i^t S_i^s=|D_i|^n \Delta_{f_i}^d=\otilde(N^n/\Delta_f^n)$).
    Since we have $\otilde(\Delta_f)$ different $f_i$ functions, we need to have $\otilde(\Delta_f)$ such instances of our
    function inversion scheme and need to query of each of them.
    This gives a total preprocessing time of $\otilde(\Delta_f\cdot (N/\Delta_f)^p)=\otilde(\Delta_f^{1-p}N^p)$,
    a total query time of $T=\otilde(\Delta_f)\cdot T_i=\otilde(\Delta_f N^{\alpha})$ and a total required space of
    $S=\otilde(\Delta_f)\cdot S_i=\otilde(\Delta_f^{1-n/s}N^{(n-\alpha t)/s})$.
    One can check that these values of $T$ and $S$ lie on the tradeoff curve
    $T^t S^s=\otilde(N^n \Delta_f^{t+s-n})$.
    It is important to note that not all values of $T,S$ that satisfy this equation are reachable using our
    parameter $\alpha$.
    Indeed, we need $T_i,S_i\geq 1$ and we use these inequalities to derive the constraints
    $0\leq \alpha\leq n/t \cdot (1-\log_N \Delta_f)$.
    In particular, we can only reach values of $T$ that vary between $T\geq \Delta_f N^0=\Delta_f$ and
    $T\leq \Delta_f N^{n/t(1-\log_{N}\Delta_f)}=N^{1+(n/t-1)(1-\log_N \Delta_f)}$.
    Since $N^{1+(n/t-1)(1-\log_N \Delta_f)}=\Omega(N)$ as $N\geq \Delta_f$ and $n\geq t$ by Lemma~\ref{lem:tradeoffexponents},
    the upper bound lies beyond the useful range of values for $T$.
    Since $S\leq N$ and $n\geq s$ by the same lemma, we have
    $T^t =N^n \Delta_f^{t+s-n}/S^s\geq (N/\Delta_f)^{n-s} \Delta_f^t\geq \Delta_f^t$, which means that the constraint $T\geq \Delta_f$
    is also always valid for nontrivial values of $S$.
    We do not enforce $T_i,S_i\leq N/(\Delta_f \log N)$ as if the value of $\alpha$ is such that either of these
    is false, we obtain a scheme that achieves either $T\geq N$ or $S\geq N$, which is outside the useful
    parameter range.

    We still need to justify that our scheme returns all preimages for all query points $y\in \Im(f)$.

    First, let us establish the success probability of the function inversion schemes on the subfunctions $f_i$.
    Each finds a preimage of $y$ with probability at least $1-(\Delta_f \log N)/N$ by hypothesis
    (because the high probability statement is with respect to the size of the domain of $f_i$, not on the domain of
    $f$).
    We can assume that $\Delta_f\leq N/\log^2 N$ (otherwise, we can do no preprocessing and use a bruteforce approach, as
    $T\geq \Delta_f$ would force $T=\omegatilde(N)$ anyway).
    Hence, if we replicate each subscheme $\Theta(\log N)$ times, we obtain a success probability
    of $1-((\Delta_f \log N)/N)^{\Theta(\log N)}\geq 1-(1/\log N)^{\Theta(\log N)}=1-N^{-\omega(1)}$ for each of them.
    The exact same comment can be made if the function inversion scheme that was used was extended using
    Theorem~\ref{thm:codomainextension}.
    Replicating them as above also gives a high probability of success in $N$.

    Now that we know that we will get a preimage for each subscheme $f_i$ if there is one, we need to show that the
    union of their answers will yield a large enough subset of $f^{-1}(y)$.
    By Lemma~\ref{lem:hasheddomainonepreimage}, the probability that $x\in f^{-1}(y)$ is never alone in some $D_i$ and
    that it is thus never returned by the inversion of $f_i$ for some $1\leq i\leq K$ is at most $2/3$.
    If we replicate the whole scheme $\Theta(\log^2 N)$ times, we obtain a probability of failure to retrieve a fixed $x$ of
    $(2/3)^{\Theta(\log^2 N)}=N^{-\omega(1)}$.
    Applying the union bound over all $O(N)$ possible values of $y$ in $\Im(f)$ yields the result.
\end{proof}

    \section{Applications}\label{sec:applications}

    Our Theorem~\ref{thm:efficientpreimagereporting} can be plugged in many constructions to retrieve existing results.
For instance, we recover the findings of Kirkpatrick et al.~\citep{kirkpatrick2026} and Carmel et al.~\citep{carmel2026}
on a data structure variant of 3SUM called preprocessed 3SUM (which is not 3SUM-indexing).
We can also recover the solution of Bibbens et al.~\citep{bibbens2026} for a text indexing problem.
As mentioned before, the solutions of Kirkpatrick et al.~\citep{kirkpatrick2026} and Bibbens et al.~\citep{bibbens2026}
involved designing their own preimage reporting scheme.

In this section, we show some applications of our main result that either give a new tool that may be of independent
interest (section~\ref{subsec:speedingupfunctioninversion}) or improve the state of the art for some
other reporting problems (section~\ref{subsec:otherapplications}) with minimal effort.

\subsection{Speeding up function inversion}\label{subsec:speedingupfunctioninversion}

We will now give a simple derived data structure that is able to retrieve a single preimage for any query point
$y\in [M]$ with a running time that depends on the number of preimages of $y$.

The idea is to build the data structure of Lemma~\ref{lem:improvedfullinversion},
but to stop after finding one preimage.
We can show that this happens with high probability after querying a fraction $\otilde(1/|f^{-1}(y)|)$ of the
subfunctions $f_i$.
When we talk of the function inversion scheme that deals with $f_i$, we actually talk about the
$\Theta(\log^2 N)$ repetitions of this scheme that are here to guarantee a high probability of success (which does not change
anything as the repetitions get absorbed by the $\otilde$ notation).

We require the following lemma.

\begin{lem}\label{lem:hasheddomainfretrievableonesi}
    With high probability and for all $y\in\Im(f)$, at least one preimage of $y$ is returned by the
    preimage reporting scheme of section~\ref{subsec:preimagereportingreduction} within $\Delta_f\log^3 N/|f^{-1}(y)|$
    inversion queries to the $f_i$ subfunctions.
\end{lem}

\begin{proof}
    Let $X$ count the number of preimages of $y$ found by the examination of $\Delta_f\log^3 N/|f^{-1}(y)|$
    subfunctions $f_i$ .
    We have $\E[X]=\mu=|f^{-1}(y)|/N\cdot \Delta_f \log^3 N/|f^{-1}(y)|\cdot N/(\Delta_f\log N)=\log^2 N$.
    Applying Lemma~\ref{lem:kwiseconcentrationbound} with $T=\mu=\log^2 N$ and
    $k=\Theta(\log N)$, we obtain that
    $\Pr[X=0]\leq \Pr[|X-\mu|\geq \mu]\leq N^{-\omega(1)}$ and at least one preimage is found with probability at least
    $1-N^{-\omega(1)}$.
    Applying the union bound over all $O(N)$ possible choices of $y\in \Im(f)$ gives the result.
\end{proof}

We can now show the following lemma.

\begin{lem}\label{lem:outputsensitiveinverters}
    Let $t,s,n,p>0$, $d\geq 0$ be constants.
    If there exists an inversion scheme that achieves the tradeoff $T^t S^s =N^n \Delta_f^d$ that preprocesses
    $f:[N]\to[N]$ in time $\otilde(N^p)$, then there exists a
    corresponding extended function inversion data structure that preprocesses $f:[N]\to[M]$ with $M=2^{\otilde(1)}$ in
    time $\otilde(\Delta_f^{1-p}N^p)$ with high probability.
    Furthermore, for all choice of $T$ and $S$ satisfying $T^t S^s =N^n \Delta_f^{t+s-n}$ and $T=\omegatilde(\Delta_f)$,
    the data structure requires space $\otilde(S)$ and finds a preimage of any query point $y\in [M]$ in time
    $\otilde(T/(1+|f^{-1}(y)|))$.
\end{lem}

\begin{proof}
    By Lemma~\ref{lem:hasheddomainfretrievableonesi}, an inverse is found with high probability after
    $\Delta \log^3 N/|f^{-1}(y)|$ queries to the function inversion scheme, which means that the space usage is
    the same as in Lemma~\ref{lem:improvedfullinversion} but the query time is only
    $\Delta \log^3 N/|f^{-1}(y)|\cdot\otilde(N^\alpha)=\otilde(\Delta_f N^{\alpha}(1+|f^{-1}(y)|)^{-1})$.

    For the same reasons as in the proof of Lemma~\ref{lem:improvedfullinversion}
    (section~\ref{subsec:preimagereportingreduction}), the above is equivalent to saying that for any choice of
    $T$ and $S$ on the curve $T^t S^s=N^n \Delta_f^{s+t-n}$ with $T=\omegatilde(\Delta_f)$, we achieve function inversion in
    time $\otilde(T (1+|f^{-1}(y)|)^{-1})$ and in space $\otilde(S)$.
\end{proof}

Applying it to Fiat-Naor's $TS^2=N^2 \Delta_f$, we find the following theorem.

\begin{thm}[Output-sensitive $TS^2=N^2 \Delta_f$]\label{thm:fiatnaoroutputsensitive}
    There exists an extended inversion data structure that, for any choice of $T$ and $S$ satisfying $TS^2=N^2 \Delta_f$
    and $T=\omegatilde(\Delta_f)$, achieves function inversion in time $\otilde(T/(1+|f^{-1}(y)|))$ and space
    $\otilde(S)$.
    Furthermore, $f:[N]\to[M]$ with $M=2^{\otilde(1)}$ is preprocessed in time $\otilde(N)$ with high probability.
\end{thm}

\subsection{{$k$SUM-indexing and Jumbled indexing with reporting}}\label{subsec:otherapplications}

3SUM-indexing is the problem where  we are given a set $A$ of $n$ integers from a universe $U=\{1,\ldots,u\}$ with
$|U|=2^{\otilde(1)}$ to preprocess to answer queries where we are given some $c\in U$, and we need to report one pair
$(a,b)\in A^2$ such that $a+b=c$, if one exists.
This problem has already been extensively studied~\citep{chung2023,dinur2026,goldstein2017,golovnev2020,kopelowitz2019}.
In 2024, Bille et al.~\citep{bille2024} considered the reporting variant of this problem (where all pairs that sum to the
query $c$ must be found).
They designed a problem-specific preimage reporting data structure to solve it and showed the following theorem.

\begin{thm}[{\authorcal{Bille, G\o rtz, Lewenstein, Pissis, Rotenberg, Steiner}~\citep{bille2024}}]\label{thm:bille3sum}
    For every $0\leq \delta\leq 1$, there is a data structure for 3SUM-indexing with reporting with
    $\otilde(n^{2-\delta/3})$ space and $\otilde((1+m)n^\delta)$ query time, where $m$ is the size of the output.
\end{thm}

We easily derive an improvement over their solution by using our Theorem~\ref{thm:efficientpreimagereporting}.
In fact, we show an efficient solution for $k$SUM-indexing with reporting, where we need to find all $(k-1)$-tuples of
elements of $A$ that sum to the given query $c\in U$.

\begin{cor}[Improved $k$SUM-indexing with reporting]\label{cor:improved3sumreporting}
    For every $0\leq \delta\leq k-1$, there is a data structure for $k$SUM-indexing with reporting with
    $\otilde(n^{k-1-\delta/2})$ space and $\otilde(\Delta n^\delta)$ query time, where $\Delta$ is the maximal size
    of the output.
    Furthermore, the data structure can be built in time $\otilde(n^{k-1})$.
\end{cor}

\begin{proof}
    Let $A[i]$ denote the $i$th element of $A$ in some arbitrary ordering.
    We consider the function $f:[n]^{k-1} \to U:(i_1,\ldots,i_{k-1})\mapsto \sum_{j=1}^{k-1}A[i_j]$.
    We preprocess it in time $\otilde(n^{k-1})$ using our
    Theorem~\ref{thm:efficientpreimagereporting} (which we can do since $[n]^{k-1}$ is in
    bijection with $[n^{k-1}]$, the correct form for the domain of the function to preprocess).
    We obtain a data structure that finds all preimages of $f$ at any given point in time $\otilde(T)$ and space
    $\otilde(S)$ for any $T,S$ on the curve $TS^2=N^2 \Delta$, with $N=n^{k-1}$.
    When given $c\in U$, the set $f^{-1}(c)$ gives the answer to the $k$SUM-indexing with
    reporting query, as we can easily convert a pair $(i_1,\ldots,i_{k-1})$ into the required
    $(A[i_1],\ldots,A[i_{k-1}])$.
    Fixing $T=\Delta n^{\delta}$ for any $0\leq \delta\leq k-1$, we find $S=n^{k-1-\delta/2}$.
\end{proof}

In the worst case and for any value of $\delta>0$, Theorem~\ref{thm:bille3sum} gives the same running time as
our Corollary~\ref{cor:improved3sumreporting} for $k=3$ but with strictly worse space: $\otilde(n^{2-\delta/3})$
instead of our $\otilde(n^{2-\delta/2})$.
Note that $\Delta$ depends on the instance $A$ (we do know that $\Delta=O(n)$ because $A$ is a set, but it could be
much smaller).
We can find its precise value in time $O(N)=O(n^2)$, which comes essentially for free in the context of the
$\otilde(N)=\otilde(n^2)$ preprocessing time of our data structure.

In their paper, Bille et al.~actually showed that a solution to 3SUM-indexing with reporting implies solutions
to many other reporting problems.
We give one such example below, in the form of the Jumbled indexing with reporting problem.

Let $S$ be a string over an alphabet $\Sigma$.
A histogram is an array $h$ of dimension $|\Sigma|$ where the entry at index $i$ gives the number of times the $i$th letter
of $\Sigma$ appears in $S$ (assuming some ordering on $\Sigma$).

In Jumbled indexing, we are given a string $S$ over $\Sigma$ to preprocess to answer queries where we are
given some histogram $P$ and need to find one substring of $S$ whose histogram is $P$ (see e.g.~\citep{amir2014}).
The reporting variant of the problem was studied by Afshani et al.~\citep{afshani2020}, who showed that if $m$ is
the size of the output, then any solution that takes time $O(n^{0.5-o(1)}+k)$ to report all substrings must consume
$\Omega(n^{2-o(1)})$ space.

\begin{cor}[Jumbled indexing with reporting, \authorcal{Bille et al.}~\citep{bille2024}]
    For every $0\leq \delta\leq 1$, there is a data structure for Jumbled indexing with reporting on alphabets
    $\Sigma$ of size $|\Sigma|=O(1)$ with $\otilde(n^{2-\delta/3})$ space and $\otilde((1+m) n^\delta)$ query time,
    where $m$ is the size of the output.
\end{cor}

Plugging our Corollary~\ref{cor:improved3sumreporting} into their reduction gives the following, which again improves
on their result for any $\delta>0$ in the worst case.

\begin{cor}
    For every $0\leq \delta\leq 2$, there is a data structure for Jumbled indexing with reporting on alphabets $\Sigma$ of size
    $|\Sigma|=O(1)$ with $\otilde(n^{2-\delta/2})$ space and $\otilde(\Delta n^\delta)$ query time, where $\Delta$ is the maximal size
    of the output.
\end{cor}

    \section*{Acknowledgements}

The main result of this work was found in the context of Mathieu Vannimmen's master's thesis, supervised by both
Jean Cardinal and John Iacono.

No AI tools were used to produce mathematical ideas, proofs or to write this document.
The authors used Claude Opus 5.5 and Claude Fable 5.1 sparingly and exclusively to proofread technical arguments.

We thank Jean Cardinal and John Iacono for many helpful discussions.

    \bibliography{refs}

    \appendix

    \section{Codomain extension}\label{sec:codomainextension}

In this appendix, we show Theorem~\ref{thm:codomainextension}.
We require the following lemma.

\begin{lem}[{\authorcal{Even, Goldreich, Luby, Nisan, Veli\v{c}kovi\'{c}}~\citep[][Theorem 2]{even1992}}]\label{lem:kwiseintersection}
    Let $X_1,\ldots,X_n$ be $k$-wise independent binary random variables.
    Then,
    \[\left|\Pr\left[\bigcap_{i=1}^n X_i=1\right]-\prod_{i=1}^n \Pr\left[X_i=1\right]\right|\leq 2^{-\Omega(k)}\]
\end{lem}

We now restate the theorem before proving it.

\begin{thm*}
    Let $t,s,n>0$, $d\geq 0$ be constants.
    If there exists a function inversion data structure that achieves the tradeoff $T^t S^s=N^n \Delta_f^d$ that
    preprocesses $f:[N]\to[N]$ in time $P$, then there exists a
    corresponding extended function inversion data structure that achieves the same tradeoff and preprocesses
    $f:[N]\to[M]$ for $M=2^{\otilde(1)}$ in time $\otilde(P)$ with high probability.
\end{thm*}

\begin{proof}
    Pick a function $h:[M]\to[N]$ from a $\lceil\log N\rceil$-wise independent hash family.
    Instead of inverting in $f:[N]\to [M]$, we invert in $h\circ f:[N]\to [N]$.
    Let $S=f([N])=\Im(f)$ be the image of $f$ and let $h_S:S\to [N]:x\mapsto h(x)$ be $h$ restricted to $S$.
    By standard balls and bins results (e.g.~\citep{celis2011}), $\Delta_{h_S}=O(\log N)$ with high probability
    (one hash $h$ is used for one function $f$, so taking $S$ to be fixed is not a problem).
    For any $z\in \Im(h\circ f)$, $(h\circ f)^{-1}(z)=\bigcup_{w\in h_S^{-1}(z)}f^{-1}(w)$, so
    $|(h\circ f)^{-1}(z)|\leq \Delta_{h\circ f}\leq \Delta_{h_S}\Delta_f=\otilde(\Delta_f)$ and the indegree of
    $f$ is preserved with high probability.

    Consider a fixed $y\in \Im(f)$.
    When given $y$, we find a preimage by $f$ by computing a preimage of $h(y)$ by $h\circ f$.
    Indeed, $x\in (h\circ f)^{-1}(h(y))$ means that $h(f(x))=h(y)$.
    The idea is that most likely, we have $f(x)=y$ but it could be that $f(x)\neq y$ and that $h$ introduces a false
    positive.
    We bound the probability that this happens.
    Let $h(y)=v$ and let $X_{z}$ be a binary random variable that is 1 if and only if $h(z)=v$ for $z\in \Im(f)$.
    Conditioned on $h(y)=v$, the variables $h(z)$ for $z\neq y$ are $(\lceil\log N\rceil-1)$-wise independent.
    By Lemma~\ref{lem:kwiseintersection} and for large enough $N$, we have:

    \[\Pr\left[\bigcap_{z\in f([N])\setminus\{y\}}X_z=0\right]\geq \left(1-\frac{1}{N}\right)^N-N^{-\Omega(1)}\geq \frac{1}{3}\]

    Hence, the probability that a collision with $y$ happens is at most $2/3$.
    Replicating the construction $\log^2 N$ times gives a probability of success for at least one of the schemes
    of $1-N^{-\omega(1)}$ and applying the union bound over all $O(N)$ possible query points $y\in \Im(f)$ yields the
    result.
\end{proof}

\end{document}